\documentclass[%
 reprint,
superscriptaddress,
 amsmath,amssymb,
 aps,
 prl,
]{revtex4-2}

\usepackage{graphicx}
\usepackage{dcolumn}
\usepackage{bm}
\usepackage{physics}
\usepackage{amssymb}
\usepackage[caption=false]{subfig}
\usepackage{color}
\usepackage{ulem}
\usepackage{cancel}
\usepackage{enumitem}
\setlist{noitemsep, topsep=0pt, parsep=0pt, partopsep=0pt}
\usepackage{comment}

\usepackage{float}
\usepackage{dsfont}

\usepackage[squaren]{SIunits}
\usepackage{hyperref}
\hypersetup{colorlinks=true,linkcolor=blue,citecolor=blue,urlcolor=blue}

\begin{document}

\preprint{APS/LLMB}

\title{Localization-landscape generalized Mott-Berezinski\u{\i} formula}

\author{Gabriel Hayoun}
\email{gabriel.hayoun@espci.psl.eu}
\affiliation{Institut Langevin, ESPCI Paris, PSL University, CNRS, Paris, France}

\author{Ilya A. Gruzberg}
\email{gruzberg.1@osu.edu}
\affiliation{Department of Physics, The Ohio State University, Columbus, Ohio 43210, USA}
 
\author{Marcel Filoche}
\email{marcel.filoche@espci.psl.eu}
\affiliation{Institut Langevin, ESPCI Paris, PSL University, CNRS, Paris, France}

\date{\today}

\begin{abstract}
We introduce a conceptual reformulation of the Mott–Berezinskiĭ (MB) theory of low-frequency AC conductivity in disordered systems based on localization landscape theory. Instead of assuming uniform localization and fixed hopping distances, transport is described through an effective potential whose geometry encodes the spatial organization and energy-dependent localization of quantum states. Using the associated Agmon metric, we define a generalized Mott scale that replaces the classical hopping length with a geometric criterion set by the disorder landscape. This framework naturally incorporates strong spatial inhomogeneity and yields the AC conductivity directly from the effective potential. The standard MB result is recovered as a limiting case. Our approach extends the conceptual foundation of MB theory to arbitrary disordered media and energies approaching the mobility edge, providing a unified description of AC transport in complex quantum materials.
\end{abstract}


\maketitle


In lightly-doped semiconductors displaying some form of quenched disorder, whether this disorder originates from a random spatial alloy composition or from the distribution of impurities, the electronic states near the Fermi energy that contribute to the conduction at low temperature can be either extended or localized, the two regimes being separated by the so-called {\it mobility edge}. On the localized side, electronic transport occurs via electron hopping between localized states, and can be modeled by a network of random resistors and capacitors~\cite{Pollak_1974_proc} and investigated through percolation theory~\cite{FishchukRudko_1980_JPhysC, Elliott_1987_AdvPhys, Hunt_2001_PhilosMagB, Shklovskii-Efros-Electronic-properties-book-1984}. The low-temperature conductivity of Weyl semimetals and weakly doped semiconductors can also be derived from a renormalization scheme developed to study localization-delocalization phase transition~\cite{Syzranov_2015_PhysRevLett}.

At low temperatures, the phonon-assisted hopping progressively leads to vanishing DC conductivity at \(T=0\), while the AC transport relies on photon-assisted hopping~\cite{TanakaFan_1963_PhysRev, BlinowskiMycielski_1964_PhysRev, BlinowskiMycielski_1965_PhysRev}. In $d$ dimensions, in the low-frequency limit, the AC conductivity~$\sigma(\omega)$ is described by the Mott-Berezinski\u{\i} (MB) formula, first proposed by Mott (with contributions by Halperin and Anderson) based on intuitive physical arguments~\cite{Mott-Conduction-I-1968, Mott-Conduction-IV-1970, Mott_1995_Book, Mott_2012}:
\begin{equation}\label{eq:MB_formula}
	\sigma(\omega) \underset{\omega\longrightarrow0}{\sim }2\pi e^2 \hbar \, \nu^2 \xi^{d+2} \omega^2 \left( \ln\frac{2\Delta_\xi}{\hbar\omega} \right)^{d+1},
\end{equation}
where $\nu$ is the density of states at the Fermi energy~$E_F$, $\xi$ is the localization length of these states, and $\Delta_\xi = (\nu \xi^d)^{-1}$ is the mean level spacing within the localization volume. In this limit, the main dependence on the frequency $\omega$ is quadratic, with a logarithmic correction whose power depends on $d$. The MB formula was rigorously derived by Berezinski\u{\i} in one dimension (1D)~\cite{Berezinskii_1973_ZhEkspTeorFiz}.

The MB formula relies on several assumptions that are critical for its proper derivation:
\begin{itemize}
    \item The Fermi energy lies in the part of the spectrum where electronic states are Anderson-localized~\cite{Anderson_1958_PhysRev}.
    \item The states contributing to $\sigma(\omega)$ are located in spatially distinct wells of the random potential.
    \item The localization length $\xi$ of the relevant states is uniform across the system.
    \item Finally, although the MB formula is expressed in any dimension, Berezinski\u{\i}'s mathematical derivation was only one-dimensional.
\end{itemize}

This formula was re-derived in 1D using various methods: the phase formalism~\cite{Abrikosov-Ryzhkin-1978, Lifshitz_1988_IntrodDisorder}, instantons and supersymmetry~\cite{Houghton_1980_PhysRevB, HaynJohn_1991_NuclPhysB}, and correlations of electronic wave functions~\cite{Gorkov_1983, Ivanov_2012_PhysRevB}. Recent work focused on an expansion in the density of potential wells and asymptotic formulas of the correlators~\cite{Kirsch_2003a} to recover the MB formula. A rigorous upper bound on \(\sigma(\omega)\) (consistent with the MB expression) was obtained in~\cite{Klein_2005_AnnMath}. Corrections to Eq.~\eqref{eq:MB_formula} and its extension to a broader range of~$\omega$ was achieved for a Gaussian white noise~\cite{Falco_2017_EPL} using the instanton approach and a modern method to compute correlation functions with functional determinants~\cite{Falco_2017_JPhysA}. The latter method was generalized to quasi-1D wires~\cite{Gruzberg-Nahum}.

The goal of the present work is to generalize the MB formula to any type of disorder, relaxing the requirement of a uniform localization length and the presence of well-defined wells in the disordered potential. We achieve these goals using the localization landscape (LL) theory.

The LL theory, introduced in 2012 by the last author and Mayboroda~\cite{FilocheMayboroda_2012_PNAS}, provides a mathematical framework to study the properties of quantum states in disordered and complex systems without having to solve the Schrödinger equation. If $\hat{H}$ is the Hamiltonian of the system, the LL~function~$u$ is defined as the solution to the associated Dirichlet problem:
\begin{equation}\label{eq:landscape_equation}
	\hat{H} u = 1 \,.
\end{equation}
A cornerstone achievement of the LL theory is the discovery that the reciprocal of the LL function, \(V_u \equiv 1/u\), acts as a true \emph{effective potential}. This emergent potential provides direct, quantitative access to the physics of localization: it delineates localization regions, predicts the integrated density of states of~\(\hat{H}\)~\cite{David_2021_AdvMath}, and determines the localization lengths of its eigenstates~\cite{Arnold_2016_PhysRevLett, Arnold_2019_SIAMJSciComput}. The LL framework has demonstrated practical impact by enabling large-scale, predictive simulations of carrier transport in highly disordered semiconductors, in particular nitride-based alloys, while achieving~${\sim}100$-fold reduction in computational cost compared to conventional quantum-mechanical approaches \cite{Filoche_2017_PRB, Piccardo_2017_PRB, Li_2017_PRB}. Recently, the~LL was used to provide a physically grounded model of variable-range hopping transport in disordered semiconductors, overcoming the simplifications of the Miller-Abrahams model~\cite{ThayilFiloche_2023_APL}.

In the following, we consider a spatial region~$\Omega$ in~$d$ dimensions in which electrons move in a time-independent (quenched) disorder potential~$V(\mathbf{r})$. We allow for the effective mass of electrons to be a position-dependent quantity $m^*_e(\vb{r})$. Then electronic wave functions $\psi$ satisfy the time-independent Schrödinger equation:
\begin{equation}
\hat{H} \psi = -\frac{\hbar^2}{2}  \mathrm{div} \left( \frac{1}{m^*_e} \vb{\nabla} \psi \right) + V  \psi = E \psi \,.
\end{equation}
We do not assume that the potential~$V$ itself exhibits clearly defined and spatially separated wells. Rather, we require this property to hold for the effective potential~$V_u$ derived from Eq.~\eqref{eq:landscape_equation}. This is a significantly weaker assumption, as $V_u$ is much smoother than $V$, a consequence of $u$ being a solution to a second-order partial differential equation. The regions surrounding the minima of $V_u$ are referred to hereafter as basins and denoted by $B_i$. These basins are defined as the areas around each minimum bounded by the level set $V_u = E$. Their extent therefore depends on the energy~$E$ under consideration.

We also assume that the basins of~$V_u$ at energy~$E_F$ are well separated and non-percolating. Since, in the lower part of the spectrum of~$\hat{H}$, electronic states are Anderson-localized within these basins~\cite{Filoche_2024_PhysRevB}, this amounts to assuming that, as in Mott’s original hypothesis, the Fermi level lies in this localized regime. However, unlike in the MB derivation, we do not require all states to share an identical localization length~$\xi$. Finally, $V_u$ is assumed to be statistically isotropic and homogeneous.

Our goal is to compute asymptotically the AC conductivity $\sigma(\omega)$ in the limit $\omega \rightarrow 0$ when the density of states per unit volume, $\nu(E)$, can be taken as a constant~$\nu$ in the range $[E_F -\hbar\omega, E_F+\hbar\omega ]$. The statistical isotropy of the medium ensures that the average conductivity is the same in all directions of space. The AC conductivity is given by the Kubo-Greenwood (KG) formula:
\begin{equation}\label{eq:KG_formula}
\sigma(\omega)\underset{\omega\longrightarrow0}{\approx} \frac{2\pi e^2 \hbar}{d} \left|\Omega\right| \nu^2 \omega^2 \, \abs{\vb{X}}_{\,\rm avg}^2 \,,
\end{equation}
where $\abs{\Omega}$ is the volume of the system and $\abs{\vb{X}}_{\rm avg}^2$ is the squared position matrix element, averaged over all possible initial and final states~\cite{Lifshitz_1988_IntrodDisorder}. The matrix element of~$x_{\alpha}$ between any initial and final states $(i,j)$ reads:
\begin{equation} \label{eq:pos_mat_elem}
	{X}_{j,i}^{(\alpha)} = \mel{\psi_j}{\hat{x}_{\alpha}}{\psi_i} = \int \psi_{j}^*(\vb{r}) \, x_{\alpha} \, \psi_{i}(\vb{r}) ~\mathrm{d} \bm{r}\,,
\end{equation}
with $\abs{\vb{X}_{j,i}}^2  = \displaystyle \sum_{\alpha=1}^{d} \abs{{X}_{j,i}^{(\alpha)}}^2$.

The first step of the derivation is to determine which pairs of states $(i,j)$ contribute to the conduction at frequency~$\omega$. To this end, we generalize the criterion introduced by Mott and Berezinski\u{\i} based on the spatial separation between states localized in two basins, $B_i$ and~$B_j$. We define the two-well Hamiltonian~$\hat{H}_{i,j}^{(2)}$ as the projection of $\hat{H}$ onto the subspace spanned by $\psi_i^{(1)}$ and $\psi_j^{(1)}$, the localized eigenstates of the one-well Hamiltonians associated with each basin. The Hamiltonian~$\hat{H}_{i,j}^{(2)}$ is thus a $2\times 2$ matrix whose off-diagonal element $t_{ij}$ is given by the overlap integral:
\begin{equation}\label{eq:tij}
	t_{ij} = t_{ji}^* \approx E_F \int_{\Omega} \psi_{j}^{(1)*}(\vb{r}) \, \psi_i^{(1)}(\vb{r}) ~{\rm d} \vb{r} \,.
\end{equation}

The two states hybridize and form bonding and anti-bonding states $\psi_+$ and $\psi_-$~\cite{Ivanov_2012_PhysRevB}, with the energy splitting
\begin{equation}\label{eq:DeltaE}
	\Delta E = E_+-E_-=\sqrt{4 t_{ij}^2 + (E_i-E_j)^2} \,.
\end{equation}
A pair of states $(i,j)$ 
contributes to $\sigma(\omega)$ only if the energy splitting satisfies the resonance condition: 
\begin{equation}
   \Delta E = \hbar \omega \,.
   \label{eq:resonance}
\end{equation}

When the overlap integrals are small, hybridization has minimal impact on the wave functions. As a result, the single-well wave function $\psi_i^{(1)}$ is nearly identical to the system’s wave function localized within the basin~$B_i$. Nevertheless, even for small $t_{ij}$, the basins $B_i$ and $B_j$ cannot contribute to the AC conductivity if they are too close to each other, since then $\Delta E$ fails to satisfy the resonance condition. Therefore, we must obtain precise estimates of the overlap $t_{ij}$, which we are going to do using the LL.

A salient point of our approach is that $V_u$, contrary to the original potential, provides in general very good estimates for the rate of decay of wave functions~\cite{Arnold_2016_PhysRevLett}. These estimates use the so-called Agmon distance~$\rho_E$ \cite{Agmon_1982_Lectures, Agmon_1985} defined as:
\begin{equation}\label{eq:Agmon_distance}
	\rho_E(\vb{r}_1, \mathbf{r}_2) = \min_{\gamma(\vb{r}_1, \vb{r}_2)} \int\limits_{\gamma} \sqrt{ \,\frac{2m}{\hbar^2}  \,\big[ \, V_u(\vb{r}) - E  \,\big]_+} ~ {\rm d}s \,,
\end{equation}
where $[x]_+ = \max(x,0)$. The minimum is computed over all possible paths between points $\vb{r_1}$ and $\vb{r_2}$, $E$ being the energy of the state considered. The path that minimizes the Agmon distance is a \emph{geodesic} of the implicit metric inside the integral. For a constant potential, this geodesic is simply a straight line between the two points, and the Agmon distance reduces to the usual term $\sqrt{2m(V - E)/\hbar^2 } \, \abs{\vb{r_2}-\vb{r_1}}$ that appears in quantum tunneling.

With the Agmon distance at hand, we can characterize the exponential decay of any localized wavefunction outside its basin, provided there are no resonances in distant basins~\cite{Arnold_2016_PhysRevLett}:
\begin{equation}\label{eq:Agmon_approx_exp}
	\psi_{i}(\vb{r}) \approx c_i \exp(-\rho_{E_i}(\vb{r}, B_i)) \qquad \text{outside } B_i \,.
\end{equation}
The detailed structure of \(\psi_i\) inside the basin is not important for what follows, but we can assume that \(\abs{\psi_i(\vb{r})} \sim c_i\) inside \(B_i\). Due to the exponential decay outside \(B_i\), the normalization constant \(c_i \approx V_i^{-1/2}\) where \(V_i\) is the localization volume occupied by the state \(\psi_i\).

Let us consider two localized states indexed by $i$ and~$j$: to ensure energy conservation for photon-assisted hopping, we require $\Delta E = \hbar\omega$, see Eq.~\eqref{eq:resonance}. Their energies are thus bounded by $E_F-\hbar\omega\leq E_{i,j}\leq E_F+\hbar\omega$ and as $\hbar\omega$ tends to zero, we have $E_i\approx E_j \approx E_F$. Agmon distances associated to both energies therefore become almost identical to the Agmon distance associated to $E_F$. From now on, we will only consider this Agmon distance, hereafter denoted~$\rho$ (removing the subscript~\(E_F\)).

Plugging Eq.~\eqref{eq:Agmon_approx_exp} into Eq.~\eqref{eq:tij} for \(t_{ij}\) leads to the computation of an integral of $\exp(-\rho(\vb{r},  \, B_i) - \rho(\vb{r}, \, B_j))$ mostly supported around the geodesic of the Agmon distance connecting basins~$B_i$ and $B_j$. This results in the estimate (see Appendix~A for details):
\begin{equation}
\label{eq:t-rho}
	t_{ij} \propto  \Delta_\xi e^{-\rho_{ij}} \rho_{ij},
\end{equation}
where $\rho_{ij}$ is the Agmon distance between basins~$B_i$ and~$B_j$. For the sake of simplicity, we retain the notation~$\Delta_\xi$ for the mean level spacing, as in Eq.~\eqref{eq:MB_formula}, even though a unique localization length~$\xi$ no longer exists.

The resonance condition of Eq.~\eqref{eq:resonance} together with Eq.~\eqref{eq:DeltaE} implies that the states \(i\) and \(j\) contribute to \(\sigma(\omega)\) only if $2 t_{ij} \leq \hbar \omega$. This imposes a lower bound $\rho_\omega$ on the Agmon distance between such states:
\begin{equation}\label{eq:Among_Mott_scale}
	\rho_{ij} \gtrsim  \rho_{\omega} = \ln\frac{2\Delta_\xi}{\hbar\omega}.
\end{equation}
Here $\rho_{\omega}$ is the LL-generalized Mott scale~\footnote{A more accurate estimate results from inverting Eq.~\eqref{eq:t-rho} using the Lambert \(W\) function: \(\rho_{ij} \approx - W_1(-t_{ij}/\Delta_\xi) \gtrsim -W_1(-\hbar \omega/2\Delta_\xi)\). For \(x \ll 1\) one can use the asymptotic expansion \(W_1(-x) = \ln(-x) - \ln(-\ln(-x))\) + o(1).}. Beyond~\(\rho_{\omega}\), the exponential decay of localized states leads to suppressed contributions to conductance. The Agmon distance is dimensionless, so the range of \(\rho\) relevant for conduction at frquency \(\omega\) is \([\rho_\omega, \rho_\omega + 1]\).

\begin{figure}[t]
\centering
\includegraphics[width=0.7\columnwidth]{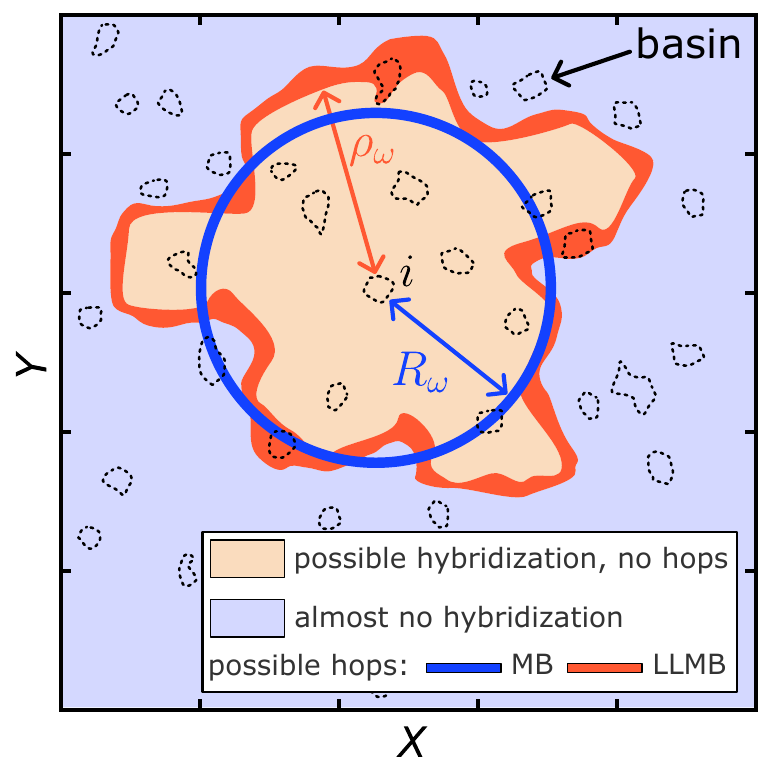}
\caption{Schematic of LL approach to AC transport. Basins \(B_i\) are displayed by black dashed lines. The orange region is such that $\rho_{\omega} \leq \rho(\vec{r}, B_i) \leq \rho_{\omega} + d\rho$, with $d\rho=1$. The width of this region is constant in Agmon's distance but not in Euclidean distance. The boundary at the Mott scale $R_{\omega}$ is the blue circle.}
\label{fig:sch_tr_mech}
\end{figure}

We can now generalize Mott's argument in terms of the Agmon distance. Figure~\ref{fig:sch_tr_mech} provides a schematic representation of the situation for a fixed value of~$\omega$. The small region in the center is the basin~$B_i$ of one localized state (of index~$i$). All points located at an Agmon distance \(\rho \in [\rho_\omega, \rho_\omega + 1]\) from this basin are plotted in orange. If another localized state lives further or closer than this distance, it cannot hybridize with state~$i$, and thus no hopping occurs between the two states. As a result, conduction essentially operates by resonant-assisted tunneling between pairs of localized states distant by about~$\rho_\omega$ in Agmon distance. If one imagines the transportation network as a graph whose vertices are the basins of the effective potential, the actual conduction network is obtained by keeping only the edges corresponding to an Agmon distance in $[\rho_\omega, \rho_\omega + 1]$.

Now that we have determined which pairs of states contribute to the conduction through the LL-generalized Mott scale~$\rho_\omega$, we are left with estimating the square modulus of the position matrix element~$\abs{\vb{X}_{i,j}}^2 $ for these pairs of states. This quantity only depends on the Euclidean distance between basins $B_i$ and $B_j$, and a rapid computation leads for these states to:
\begin{equation}
    \abs{\vb{X}_{i,j}}^2\approx 
    \abs{\vb{r}_i - \vb{r}_j}^2/4 \,,
\end{equation}
(see Appendix~B for details). For states that do not contribute to the conductivity, $\abs{\vb{X}_{i,j}}^2\approx 0$.

We define $\Sigma_\omega$ as the set of state indices~$i$ satisfying $\abs{E_i - E_F} \le \hbar \omega$. Other states cannot participate in the conduction. Let us compute the average distribution of Euclidean distance between pairs of states  $(i,j) \in \Sigma_\omega^2$ separated by the Agmon distance $ \rho_{ij}\in[\rho,\rho + \mathrm{d}\rho]$ and the Euclidean distance $r_{ij}=\abs{\vb{r}_i - \vb{r}_j} \in [r, r+\mathrm{d}r]$. This distribution is defined as follows:
\begin{align}
	f(r,\rho) \, \frac{\mathrm{d}r \,\mathrm{d}\rho}{\abs{\Omega}} = 
\ev{\mathds{1}_{[r,r+\mathrm{d}r]}(r_{ij}) \times \mathds{1}_{[\rho,\rho+\mathrm{d}\rho]}(\rho_{ij})}_{i,j\,\in \,\Sigma_\omega},
\end{align}
where $\mathds{1}_A$ is the characteristic function of a set $A$, and $f(r,\rho)\,\mathrm{d}r \,\mathrm{d}\rho$ is the average volume filled by the final states contributing to conductivity.

The next step is to determine $\abs{\vb{X}}_{\,\rm avg}^2$ which is obtained by averaging $r^2/4$ with the above distribution:
\begin{equation}\label{eq:X_general}
	\abs{\vb{X}}_{\rm avg}^2 =  \frac{1}{4 \abs{\Omega}} \int_{r=0}^{+\infty} r^{2} f(r, \rho) \,{\rm d}\rho \,{\rm d}r.
\end{equation}
Recalling that the relevant range \(\rho \in [\rho_\omega, \rho_\omega + 1]\), we can rewrite:
\begin{align}\label{eq:X_moment}
	\abs{\vb{X}}_{\rm avg}^2 &=  \frac{1}{4 \abs{\Omega}}
\int_0^\infty
\! r^{2} f(r, \rho_\omega) ~{\rm d}r 
    = \frac{1}{4 \abs{\Omega}} \mathbb{E}_\omega(X^{2}).
\end{align}
$\mathbb{E}_\omega(X^{2})$ is the second moment in $r$ of the distribution $f$ at the LL-generalized Mott scale $\rho_{\omega}$. In the view of Eq.~(3') in~\cite[p.~854]{Gorkov_1983}, $f/\abs{\Omega}$ is an approximation of the two-point correlator given in Eq.~(4) in the same paper. In the LL framework however, only the dominant contributions to this correlator, corresponding to $\rho_{ij} \in [\rho_\omega,\rho_{\omega} + 1]$, are kept. In addition, $f$ is easier to compute than the correlator as it doesn't require a priori knowledge of the wave functions.

Substituting Eq.~\eqref{eq:X_moment} into the KG formula of Eq.~\eqref{eq:KG_formula} finally leads to the LL-generalized MB formula:
\begin{equation}\label{eq:LLeMB_formula}
\sigma\left(\omega\right)\underset{\omega\longrightarrow0}{\approx} \frac{\pi e^2 \hbar}{2d}\, \nu^2\,\omega^2 \, \mathbb{E}_\omega(X^{2}).
\end{equation}
The LL enters this formula through the LL-generalized Mott scale~$\rho_{\omega}$ that selects participating states.

Computing the conductivity of a system therefore requires to determine $f(r, \rho_{\omega})$ and then $\mathbb{E}_\omega(X^{2})$, according to the following workflow: 
\begin{align*}
    V &\longrightarrow u \longrightarrow 
        V_u \longrightarrow \left\{B_i\right\} \longrightarrow 
        \Sigma_\omega \\
        &\longrightarrow f(r,\rho_{\omega}) \longrightarrow \mathbb{E}_\omega(X^{2} ) \longrightarrow \sigma (\omega).
\end{align*}
This approach generalizes the MB derivation, allowing to investigate systems out of the scope of the MB formula.

As a consistency check, let us recover the MB formula under Mott's assumptions. In Mott's derivation, the Euclidean distance between pairs of states that participate to conduction typically belongs to \([R_{\omega},R_{\omega}+\xi]\), where $R_{\omega}$ is the Mott scale~\cite{Mott_1967_AdvPhys, Mott_1995_Book, Mott_2012} defined as:
\begin{equation}\label{eq:Mott_scale}
	R_{\omega} = \xi  \ln\frac{2\Delta_\xi}{\hbar \omega}.
\end{equation}
It can be retrieved from the LL-generalized Mott scale if one assumes that all states at a given energy share the same localization length. In this case, the Agmon distance between two states is simply $\rho_{ij} = r_{ij}/\xi$, and
\begin{equation}
    \mathds{1}_{[\rho_\omega,\rho_\omega+1]}(\rho_{ij}) = \mathds{1}_{[R_\omega,R_\omega+\xi]}(r_{ij}).
\end{equation}
This leads to:
\begin{align}
	f_{\text{Mott}}(r, \rho_{\omega})\,\mathrm{d}r = 
	\begin{cases} 
		S_d \,r^{d-1}\,\mathrm{d}r , &\text{if }R_{\omega} \leq r \leq R_{\omega} + \xi\\
		0, & \text{otherwise}
	\end{cases},
\end{align}
where $S_d$ is the area of the $d$-dimensional unit sphere. Indeed, averaging over all pairs of states in $\Sigma_\omega^2$ yields the volume occupied by the final states, which corresponds to the volume of the shell between $r$ and $r+\mathrm{d}r$. Since the basins are considered to be well separated, then $R_{\omega}\gg\xi$ and:
\begin{align}
    \mathbb{E}_\omega(X^{2})\approx S_d \, R_{\omega}^{\: d+1} \xi= S_d \, \xi^{d+2} \left(\ln \frac{2 \Delta_\xi}{\hbar\omega}\right)^{d+1}.
\end{align}
We recover here the MB formula from~Ref.~\cite{Kirsch_2003a} with an additional factor $2$ due to the inclusion of spinful electrons in our model.

Let us emphasize once more differences between the conventional Mott picture and our treatment using the~ LL. In the MB picture electronic conduction occurs by hopping between localized states separated by the Euclidean distance $R_\omega$, regardless of the details of the potential in between. The points at a distance equal to the Mott scale from a given localized state form a $(d-1)$-sphere, as depicted in blue in Fig.~\ref{fig:sch_tr_mech}. Yet, two localized states at distance $R_\omega$ from each other but separated by a very large potential barrier wouldn't be able to hybridize, and thus to contribute to conductivity. Conversely, a pair of states closer to each other than~$R_{\omega}$ can participate in transport if a large potential barrier separates them. 

These configurations, not accounted for precisely in the MB derivation, are handled in a very general way by the LL through the Agmon distance which deforms the space geometry to account for the shape of the potential. As a result the region of ``active states'' in not a Euclidean sphere but its deformed version shown in orange in Fig.~\ref{fig:sch_tr_mech}. In addition to being more general, the LL-generalized Mott scale also allows us to consider specific realizations of disorder instead of disorder averages.
    
In conclusion, the LL approach extends the MB formula to systems with statistically isotropic and homogeneous disordered effective potentials, accommodating spatially varying localization lengths. It introduces an extended Mott scale which captures fluctuations of the effective potential. Hops between states occur at Agmon distances defined by this LL-generalized Mott scale, ensuring both hybridization and resonance. In this framework, conductivity is governed by the second moment of the distribution of Euclidean distances between wells separated by approximately~$\rho_{\omega}$ in the LL-based Agmon metric.

In this work, we relaxed two fundamental assumptions of the original MB formula: the uniformity of the localization length and the requirement for well-defined wells. Future developments could involve abandoning the assumption of a constant density of states, leveraging LL-based approximations~\cite{Arnold_2016_PhysRevLett}. It could be interesting to look for specific forms of disorder that, through anomalous density of states behavior or non conventional Agmon distance, would alter the frequency dependence in the Mott-Berezinski\u{\i} equation, opening the possibility of tailoring conductivity at the nanoscale in such media.

\section*{Acknowledgments}

M. Filoche and G. Hayoun are supported by the project Localization of Waves of the Simons Foundation (Grant No. 1027116, M.F.).

\appendix

\subsection{Appendix A: Computing the transition amplitude}
\label{app:transition_amp}

We assess the transition amplitude between two ground states $\psi_i^{(1)}$ and $\psi_j^{(1)}$ of the one-well Hamiltonians $H_i^{(1)}$ and $H_j^{(1)}$, localized in the basins $B_i$ and $B_j$ with energies $E_i$ and $E_j$, both energies being approximately equal to $E_F$~\cite{Kirsch_2003a}:
\begin{equation}
	t_{ij} = E_F \int_{\Omega} \psi_{j}^{(1)*}(\vb{r}) \, \psi_i^{(1)}(\vb{r}) ~{\rm d} \vb{r} \,.
\end{equation}
Replacing both wave functions outside their basins by their approximations obtained from the LL, see Eq~\eqref{eq:Agmon_approx_exp}, leads to an expression which can be split into three contributions: one per basin and one in the domain complementary to both basins. This reads:
\begin{align}\label{Eq:tij}
	t_{ij} \approx E_F \, \bigg(
	& c_j \int_{B_i} \psi_i(\vb{r}) \, e^{-\rho(B_j, \vb{r})}~{\rm d} \vb{r} \nonumber\\
	~+~& c_i \int_{B_j} \psi_j(\vb{r}) \, e^{-\rho(B_i, \vb{r})}~{\rm d} \vb{r} \nonumber\\
	~+~& c_i \, c_j \int_{\Omega\setminus(B_i \cup B_j)} e^{-\rho(B_i, \vb{r})-\rho(B_j, \vb{r})} ~{\rm d} \vb{r}
	\bigg) \,.
\end{align}   
Let us evaluate the last contribution $I_{ij}$ defined as
\begin{equation}
	I_{ij} =  \int_{\Omega\setminus(B_i \cup B_j)} e^{-\rho(B_i, \vb{r})-\rho(B_j, \vb{r})} ~{\rm d} \vb{r} \,.
\end{equation}
To do so, we use the Laplace method, first introduced by Laplace in 1774 for one-dimensional integrals~\cite{Laplace_1774}, then extended to multivariate ones~\cite{Wong_2001_SIAM, Lapinski_2019}. This method is the real counterpart of the stationary phase method~\cite{Arnold_1981, Duistermaat_2011} used to approximate complex integrals of the form:
\begin{equation}
	I(t)  = \int_A e^{-f(\vb{r},t)} \, g(\vb{r},t) ~{\rm d} \vb{r}\,,
\end{equation}
where $A$ is a subdomain of $\mathbb{R}^d$. In our case $A = \Omega\setminus(B_i \cup B_j)$,  $g(\vb{r}, t)=1$, and $f(\vb{r}, t) = \rho(B_i, \vb{r})+\rho(B_j, \vb{r})$. Both functions $f$ and $g$ are time-independent.
	
By triangular inequality of the Agmon distance, the value of $f$ is always larger than the Agmon distance~$\rho_{ij}$ between $B_i$ and $B_j$. By definition, $f$ is equal to $\rho_{ij}$ on the geodesic of the Agmon distance connecting $B_i$ to $B_j$:
\begin{equation}
\forall \vb{r} \in  \Gamma_{ij}\,, \:f(\vb{r}) = \rho(B_i, \vb{r})+\rho(B_j, \vb{r}) = \rho_{ij} \,.
\end{equation}
We assume here that $\Gamma_{ij}$ is unique everywhere the Agmon metric is non degenerate, i.e., where $1/u(\vb{r})$ is larger than~$E_F$. We will see later that our demonstration can be extended to the case of multiple geodesics. The parts of $\Gamma_{ij}$ where the Agmon distance is degenerate ($1/u<E_F$) correspond to a union of regions~$\{\tilde{B}_k\}$. It is easy to assess the contribution of each of these regions to the coupling:
\begin{align}
	\int_{\tilde{B}_k} e^{-f(\vb{r})} ~{\rm d} \vb{r} = e^{-\rho_{ij}} \int_{\tilde{B}_k}  ~{\rm d} \vb{r} = e^{-\rho_{ij}} |\tilde{B}_k | \,,
\end{align}
where $|\tilde{B}_k|$ denotes the volume of the region~$\tilde{B}_k$.

One now focus on the complementary region, the part of the geodesic where the Agmon distance is non degenerate, hereafter called~$\Gamma_1$. Starting from any point~$\vb{s}$ of $\Gamma_1$, moving orthogonally to $\Gamma_1$, i.e., in the hyperplane $\mathcal{P}(\vb{s})$ orthogonal to the local tangent to $\Gamma_1$ at $\vb{s}$, increases the value of $f$. We can therefore expand $f$ around a point~$\vb{s}$ of $\Gamma_1$, $\vb{z}$ belonging to $\mathcal{P}(\vb{s})$:
\begin{equation}\label{eq:expansion_f}
	f(\vb{s} + \vb{z}) = \rho_{ij} + \frac{1}{2} \,\vb{z}^T H_\perp(\vb{s}) \, \vb{z} + o(\norm{\vb{z}}^2) \,,
\end{equation}
where $H_\perp(\vb{s})$ denotes the Hessian of $f$ restricted to $\mathcal{P}(\vb{s})$. By definition of the geodesic,
\begin{equation}
\forall \vb{s} \in \Gamma_1 \,,\: H_\perp(\vb{s}) > 0 \,.
\end{equation}
Using~\cite[3.1.1]{Ludewig_2017aa} (also used in \cite{Ludewig_2018}) allows us then to approximate $I_{ij}$ at the first non vanishing order:
\begin{align}\label{eq:approx_Iij}
	I_{ij}  \approx e^{-\rho_{ij}} \left[ \int_{\Gamma_1} \frac{(2\pi)^{\frac{d-1}{2}}}{\sqrt{\det(H_\perp(\vb{s}))}}  \,{\rm d} \vb{s} \,+\, \sum_k |\tilde{B}_k| \right]\,.
\end{align}
The narrower is the function~$f$ around its minimal path, the better the approximation. The quantity $1/\sqrt{\det(H_\perp(\vb{s}))}$ that appears in the denominator comes from neglecting the high-order contribution in Eq.~\eqref{eq:expansion_f}. We are left in this case with integrating a Gaussian function over the entire local hyperplane. Interestingly, $1/\sqrt{\det(H_\perp(\vb{s}))} = 1/\sqrt{\prod_\ell \lambda_\ell}$, where $\{\lambda_\ell\}$ are the eigenvalues of the Hessian, i.e., the curvature of the Agmon distance along the various transverse directions. The integrand in the first term of Eq.~\eqref{eq:approx_Iij} can therefore be interpreted as a local effective cross section. A large curvature corresponds to a small cross section, while a small curvature leads to a large cross section.

Once integrated along $\Gamma_1$, this cross section gives the effective volume of a $d$-dimensional tube connecting the various basins $\tilde{B}_k$. Consequently, one can regroup all terms inside the same expression,
\begin{equation}
	I_{ij} = e^{-\,\rho_{ij}} \,V_{ij} \, ,
\end{equation}
where $V_{ij}$ denotes the total effective volume of a tube surrounding the geodesic that significantly contributes to the coupling between localized states in basins $B_i$ and~$B_j$. If several geodesics connect both basins, the same method can be extended, summing over all volumes of each geodesic.

Returning to Eq.~\eqref{Eq:tij}, we estimate the remaining two terms. The Agmon distances involved are
\begin{align}
    \rho(B_i,\vb{r} \in B_j) = \rho(B_j,\vb{r} \in B_i) = \rho(B_i,B_j) = \rho_{ij} \,.
\end{align}
Using the estimate \(\psi_i \sim c_i\) inside \(B_i\), we get that
\begin{align}
    c_j \int_{B_i} \psi_i(\vb{r}) \, e^{-\rho(B_j, \vb{r})} {\rm d} \vb{r} \sim c_i c_j e^{-\rho_{ij}} \abs{B_i} \,.
\end{align}
The transition amplitude can now be written as
\begin{equation}
	t_{ij} \approx e^{-\rho_{ij}} \, E_F \, c_i c_j \big(V_{ij} + \abs{B_i} + \abs{B_j} \big) \,.
\end{equation}
Recall that the normalization constant~$c_i \sim V_i^{-1/2}$ with~$V_i$ the volume occupied by the state $\psi_i^{(1)}$. The effective potential being statistically homogeneous, the volumes occupied by localized states at an energy close to the Fermi energy are comparable. Therefore $c_i \, c_j \approx 1/V_F$ with $V_F$ the volume occupied by a localized state at energy~$E_F$. This volume can be expressed in terms of the mean level spacing $\Delta_\xi$,
\begin{equation}
	\Delta_\xi = \frac{1}{V_F \, \nu}, \quad \text{ hence } \quad c_i \, c_j \approx \Delta_\xi \, \nu\,.
\end{equation}
The transition amplitude now reads:
\begin{equation}\label{Eq:tij_approx}
	t_{ij} \approx  \Delta_\xi e^{-\rho_{ij}} \nu E_F \big(V_{ij} + \abs{B_i} + \abs{B_j} \big) \,.
\end{equation}
The product \(\nu E_F = n\) is, roughly, the density of carriers, and then \(N_i = n \abs{B_i}\) represents the number of carriers inside the basin \(B_i\), which is independent on the frequency~\(\omega\). On the other hand, the number of carriers inside the tube surrounding the Agmon geodesic \(N_{ij} = n V_{ij} \propto \rho_{ij}\). For the states of interest in this paper, the Agmon distance \(\rho_{ij} \gg 1\). Thus \(N_{ij} \gg N_i, N_j\), and the final estimate of the transition amplitude is
\begin{align}
    t_{ij} \propto \Delta_\xi \rho_{ij} e^{-\rho_{ij}} \,,
\end{align}
with dimensionless prefactor independent of~\(\omega\).

In 2D and approximately in 3D, one can go one step further and assess the value of the local cross section along the geodesic. To that end, one needs to estimate ${\rm det}(H_\perp(\vb{s}))$, see Eq.~\eqref{eq:approx_Iij}. This is done by plugging the estimate from Eq.~\eqref{eq:Agmon_approx_exp} into the Schrödinger equation:
\begin{equation}
    \frac{\hbar^2}{2m}\Delta\rho - \frac{\hbar^2}{2m}\abs{\nabla\rho}^2 +V \approx E\, .
\end{equation}
The eikonal equation satisfied by the Agmon distance $\abs{\nabla\rho}^2 = \frac{2m}{\hbar^2}\left(\frac{1}{u}-E\right)_+$ yields the value of $\Delta \rho$ wherever $1/u \ge E$:
\begin{equation}
    \frac{\hbar^2}{2m}\Delta\rho \approx \frac{1}{u} - V \quad \text{hence} \quad \Delta \rho \approx -\frac{\Delta u}{u}\,.
\end{equation}
The eigenvalue of the aforementioned Hessian $H_\perp$ vanishes along the direction of the geodesic. In 2D, this gives $\Delta\rho(\vb{s}) = \lambda_{\parallel} + \lambda_\perp = \lambda_\perp = \det(H_\perp(\vb{s}))$, hence:
\begin{equation}
   \det(H_\perp(\vb{s}))\approx -\frac{\Delta u(\vb{s})}{u(\vb{s})} \,. 
\end{equation}
In 3D, $\Delta\rho(\vb{s}) =  \lambda_{\perp,1} + \lambda_{\perp,2}$. Moreover, $\lambda_{\perp,1} \approx \lambda_{\perp,2}\approx \lambda_\perp$ due to the stastitical isotropy. This yields:
\begin{equation}
   \det(H_\perp(\vb{s})) \approx \lambda_\perp^2 \approx \frac{1}{4} \left(\frac{\Delta u(\vb{s})}{u(\vb{s})}\right)^2 \,. 
\end{equation}

\subsection{Appendix B: Derivation of the matrix element}
\label{app:X_ij}

One can express the wavefunctions, $\psi_{\pm}$ of the two-well Hamiltonian in terms of $\psi_{i}^{(1)}$ and $\psi_{j}^{(1)}$~\cite{Kirsch_2003a}:
\begin{align}
	&\psi_+ = \cos(\theta) \, \psi_{i}^{(1)} + \sin(\theta) \, \psi_{j}^{(1)}\, ,\\
	&\psi_- = -\sin(\theta) \, \psi_{i}^{(1)} + \cos(\theta) \, \psi_{j}^{(1)} \, ,
\end{align}
where $\theta$ is a mixing angle such that $\tan(\theta) = \frac{t_{ij}}{\delta + \sqrt{t_{ij}^2 + \delta^2}}$ with $ \delta =\frac{E_i-E_j}{2}$.
For instance, the matrix element along direction $Ox$ reads:
\begin{align}
	\vb{X}_{i,j}^{(x)} = &\int \psi_{+}^*(\vb{r}) \, x \, \psi_{-}(\vb{r})  ~\mathrm{d} \bm{r} =  \frac{t_{ij}}{2\sqrt{t_{ij}^2 + \delta^2}} (x_j-x_i) \nonumber \\
	& \quad + \frac{\delta}{\sqrt{t_{ij}^2 + \delta^2}}\int \psi_{i}^{(1)}(\vb{r}) \, x \, \psi_{j}^{(1)}(\vb{r})  ~\mathrm{d} \bm{r} \,.
\end{align}
For states that hybridize, $t_{ij} \gg \delta$ and the second term is negligible. For states that do not hybridize, the overlap is very small and the contribution of theses states to the average is negligible. In a statistically isotropic and homogeneous medium, the square matrix element reads:
\begin{equation}
	\abs{\vb{X}_{i,j}}^2  = \sum_{\alpha=1}^{d}\abs{\vb{X}_{i,j}^{(\alpha)}}^2 \,,
\end{equation}
where the sum runs over all directions of space. Consequently, $\abs{\vb{X}_{i,j}}^2\approx \frac{\abs{\vb{r}_i - \vb{r}_j}^2}{4}$ between states that contribute to conduction whereas $\abs{\vb{X}_{i,j}}^2\approx 0$ between states that do not.

\bibliography{LLMB.bib}

\end{document}